%% file: main.tex
\documentclass[conference]{IEEEtran}
\IEEEoverridecommandlockouts

\input{macros}

\begin{document}

\input{glossary}

\title{Optimized Detection with Analog Beamforming for Monostatic Integrated Sensing and Communication
}

\input{content/authors}

\maketitle

\IEEEpubidadjcol

\input{content/abstract}

\glsresetall

\section{Introduction}\label{sec:intro}
\input{content/intro}

\section{Problem Formulation}\label{sec:problem}
\input{content/problem}

\section{Numerical Results}\label{sec:results}
\input{content/results}

\section{Conclusions}\label{sec:conclusions}
\input{content/conclusions}

\section*{Acknowledgment}
\input{content/ack}

\bibliographystyle{IEEEtranN}
\bibliography{IEEEabrv,references}

\end{document}

%% file: macros.tex
\usepackage{subfigure}
\usepackage{smartdiagram}
\usepackage[numbers,compress]{natbib}
\usepackage{amsmath,amssymb,amsfonts}
\usepackage{mathrsfs}
\usepackage{mathtools}
\usepackage{graphicx}
\usepackage{caption}
\usepackage{textcomp}
\usepackage{xcolor}
\usepackage{booktabs}
\usepackage[retain-explicit-plus,qualifier-mode=text]{siunitx}
\usepackage{soul}
\usepackage{xcolor}
\usepackage{comment}
\usepackage[printonlyused,nohyperlinks, nolist]{acronym}
\usepackage[shortcuts,acronym,automake=false]{glossaries}
\usepackage{multirow}

\usepackage{enumitem}

\usepackage{amsmath}
\usepackage{amssymb}

\usepackage{stfloats}

\usepackage{url}

\usepackage{multirow}

\usepackage[hidelinks,bookmarks=false]{hyperref}

\usepackage[printonlyused,nohyperlinks,nolist]{acronym}
\usepackage{algorithm}
\usepackage{algpseudocode}
\usepackage[capitalise]{cleveref}

\usepackage{dsfont}

\usepackage{siunitx}
\DeclareSIUnit{\belmilliwatt}{Bm}
\DeclareSIUnit{\dBm}{\deci\belmilliwatt}
\usepackage{color}

\DeclareUnicodeCharacter{2212}{-}

\def\BibTeX{{\rm B\kern-.05em{\sc i\kern-.025em b}\kern-.08em
    T\kern-.1667em\lower.7ex\hbox{E}\kern-.125emX}}

\def\a{{\mathbf a}}
\def\aT{\a_{\Mt}}
\def\aR{\a_{\Mr}}
\def\c{{\mathbf c}}

\def\s{{\mathbf s}}

\def\e{{\mathbf e}}
\def\g{{\mathbf g}}
\def\v{{\mathbf v}}
\def\x{{\mathbf x}}

\def\h{{\mathbf h}}
\def\w{{\mathbf w}}

\def\gm{\g_{m}}

\def\wr{w_{\textrm r}}

\def\qstar{{q_{\star}}}
\def\wstar{{\w_{\star}}}
\def\cstar{\c_{\star}}

\def\yr{y_\mathrm{r}}

\def\sigr{\sigma_\mathrm{r}}

\def\Null{{\mathbf 0}}

\def\Mt{{M_\mathrm{t}}}
\def\Mr{{M_\mathrm{r}}}

\def\A{{\mathbf A}}
\def\B{{\mathbf B}}

\def\Hsi{{\mathbf H}_{\mathrm{si}}}
\def\bHsi{\Bar{\mathbf H}_{\mathrm{si}}}
\def\Hu{{\mathbf H}_{\mathrm{U}}}
\def\Hv{{\mathbf H}_{\mathrm{V}}}
\def\bHr{\Bar{\mathbf H}_{\mathrm{r}}}
\def\Hr{{\mathbf H}_{\mathrm{r}}}
\def\X{{\mathbf X}}
\def\U{{\mathbf U}}
\def\V{{\mathbf V}}
\def\C{{\mathbf C}}
\def\S{{\mathbf S}}

\def\I{{\mathbf I}}

\def\P{{\mathbf P}}
\def\W{{\mathbf W}}

\def\chisq{\chi^2_2}

\def\Hnull{\mathrm{H}_0}
\def\Halt{\mathrm{H}_1}
\def\Pd{P_{\mathrm{D}}}
\def\Pfa{P_{\mathrm{FA}}}

\def\Pt{P_{\mathrm{t}}}
\def\Pant{P_{\mathrm{ant}}}
\def\Psat{P_{\mathrm{sat}}}

\def\THs{{\Theta}_{\mathrm{S}}}
\def\THl{{\Theta}_{\mathrm{L}}}
\def\THe{{\Theta}_{\mathrm{E}}}
\def\gams{{\gamma}_{\mathrm{s}}}

\def\CN{\mathcal{CN}}

\def\hilbert{{\mathcal H}}

\def\RR{{\mathbb R}}
\def\Rp{{\RR_+}}

\def\CC{{\mathbb C}}
\def\EE{{\mathbb E}}
\def\VV{{\mathbb V}}
\def\HH{{\mathbb H}}
\def\NN{{\mathbb N}}
\def\ZZ{{\mathbb Z}}

\def\trace{\mathrm{tr}}

\def\maximize{\mathrm{maximize}}
\def\argmin{\mathrm{arg\ min}}
\def\argmax{\mathrm{arg\ max}}

\def\st{\mathrm{s.t.}}
\def\rank{\mathrm{rank}}
\def\diag{\mathrm{diag}}

\def\setA{\mathcal{A}}
\def\setB{\mathcal{B}}
\def\setC{\mathcal{C}}

\def\setF{\mathcal{F}}
\def\setH{\mathcal{H}}
\def\setI{\mathcal{I}}
\def\setK{\mathcal{K}}
\def\setL{\mathcal{L}}

\def\setN{\mathcal{N}}
\def\setP{\mathcal{P}}

\def\setR{\mathcal{R}}
\def\setS{\mathcal{S}}

\def\setU{\mathcal{U}}
\def\setV{\mathcal{V}}
\def\setY{\mathcal{Y}}

\def\Ftr{\setF_\mathrm{t,r}}
\def\Fj{\setF_\mathrm{j}}

%% file: glossary.tex
\newacronym{isac}{ISAC}{integrated sensing and communication}
\newacronym{music}{MUSIC}{multiple signal classification}
\newacronym{tx}{TX}{transmission}
\newacronym{rx}{RX}{reception}
\newacronym{mse}{MSE}{mean squared error}
\newacronym{ula}{ULA}{uniform linear array}
\newacronym{snr}{SNR}{signal-to-noise ratio}
\newacronym{glrt}{GLRT}{generalized likelihood ratio test}
\newacronym{sll}{SLL}{sidelobe level}
\newacronym{si}{SI}{self-interference}

\newacronym{aps}{APS}{angular power spectrum}

\newacronym{adc}{ADC}{analog-to-digital converter}
\newacronym{spocs}{SPOCS}{superiorized projections onto convex sets}

\newacronym{roc}{ROC}{receiver operating characteristic}

\newacronym{ofdm}{OFDM}{orthogonal frequency-division multiplexing}
\newacronym{qpsk}{QPSK}{quadrature phase shift keying}

%% file: content/authors.tex
\author{\IEEEauthorblockN{Rodrigo Hernang\'{o}mez\IEEEauthorrefmark{1},
Jochen Fink\IEEEauthorrefmark{1}, Renato L.G. Cavalcante\IEEEauthorrefmark{1}, Zoran Utkovski\IEEEauthorrefmark{1}, S{\l}awomir Sta\'{n}czak\IEEEauthorrefmark{1}\IEEEauthorrefmark{2}}

\IEEEauthorblockA{\IEEEauthorrefmark{1}Fraunhofer Heinrich Hertz Institute,  Germany, \{firstname.lastname\}@hhi.fraunhofer.de}
\IEEEauthorblockA{\IEEEauthorrefmark{2}Network Information Theory Group,
Technische Universit\"{a}t Berlin, Germany}
}

\IEEEpubid{%
    \begin{minipage}{\columnwidth}
    \vspace{1em}
    \copyright 2024 IEEE. Personal use of this material is permitted. Permission from IEEE must be obtained for all other uses, in any current or future media, including reprinting / republishing this material for advertising or promotional purposes, creating new collective works, for resale or redistribution to servers or lists, or reuse of any copyrighted component of this work in other works.
    \end{minipage}%
    \hspace{1em}%
    \begin{minipage}{\columnwidth}
        $ $
    \end{minipage}
}

%% file: content/abstract.tex
\begin{abstract}
In this paper, we formalize an optimization framework for analog beamforming in the context of monostatic \gls{isac},
where we also address the problem of self-interference in the analog domain.
As a result, we derive semidefinite programs to approach detection-optimal transmit and receive beamformers, and we devise a superiorized iterative projection algorithm to approximate them. Our simulations show that this approach outperforms the detection performance of well-known design techniques for \gls{isac} beamforming, while it achieves satisfactory self-interference suppression.
\end{abstract}

\begin{IEEEkeywords}
\Acrlong{isac}, beamforming, self-interference suppression, target detection, projections onto convex sets.
\end{IEEEkeywords}

%% file: content/intro.tex
\Gls{isac} has been identified as one key component of next-generation mobile networks, with application areas ranging from health and smart homes to vehicular networks and industry~\cite{liu_integrated_2022}. Indeed, the current trend to automatize industrial and automotive systems often leads to a growing number of radiofrequency hardware and software components to satisfy their sensing and communication needs. In this context, the development of small-form-factor and cost-efficient \gls{isac} solutions is particularly interesting to temper down the increased cost, size, and energy consumption that the added functionalities may bring along.

Monostatic \gls{isac}, i.e., with co-located \gls{tx} and \gls{rx} antennas, emerges as the best modality from those discussed in the literature to achieve a compact and efficient integration in standalone devices. Here, the inherent synchronization between \gls{tx} and \gls{rx} and the full knowledge of the transmitted signal are seen as its main advantages against the bistatic counterpart~\cite{liu_integrated_2022,liu_joint_2023}. 
On the other hand, the desired continuous transmission of data rules out pulse radar techniques and gives rise to \gls{si}, which hampers monostatic sensing.

A strong \acrlong{si} requires the cancellation of the unwanted signal via digital signal processing techniques, such as moving target indication or adaptive filtering. Moreover, the presence of \acrlong{si} can also cause problems to the analog circuits, e.g., \gls{rx}-amplifier saturation or an insufficient dynamic range at the \gls{adc}~\cite{askar_active_2014}.
Practical systems, typically based on \gls{ofdm}  in the millimeter-wave band \cite{baquero_barneto_millimeter-wave_2022,wild_6g_2023}, often circumvent \acrlong{si} through a moderate \gls{tx}-\gls{rx} separation ($\approx$\SI{50}{\centi\meter}). Nevertheless, this leads to physically large devices that may be unpractical in the mentioned setups.

As an alternative to suppress \acrlong{si},
full-duplex communication schemes~\cite{askar_active_2014}
have recently inspired the exploration of beamforming
in its different architectures: digital~\cite{he_full-duplex_2023}, hybrid~\cite{liu_joint_2023}, or analog~\cite{liu_full-duplex_2023}.
Specifically, \citet{liu_full-duplex_2023} design \gls{tx} and \gls{rx} analog beamformers such that the \acrlong{si} is projected into their null space while allocating beams for sensing and communication.
In doing so, the goal is to reduce the \gls{sll}.

Analog beamforming is particularly interesting due to its ability
to address \acrlong{si} in the analog domain without dedicated canceller circuits~\cite{liu_joint_2023,he_full-duplex_2023,liu_full-duplex_2023}; however, it comes at the expense of not being able to estimate the angle via array-processing algorithms, such as \gls{music}~\cite{liu_integrated_2022}.
Against this background, we have identified angle-selective target detection as the key performance indicator of analog-beamforming design for sensing. Thus, we build upon the heuristic method in~\cite{liu_full-duplex_2023}
to formalize a framework for optimal detection given the constraints imposed by communication, \acrlong{si}, and power. We propose parallel \gls{tx} and \gls{rx} optimization to approximate the optimal solution using a superiorized projection algorithm.
This technique, which has been previously applied to wireless problems such as
MIMO detection~\cite{fink_set-theoretic_2022} or
multicast beamforming~\cite{fink_multicast_2019,fink_multi-group_2021},
follows the superiorization principle~\cite{censor2010perturbation} by adding bounded perturbations to an iterative projection algorithm~\cite{theodoridis_adaptive_2011}.
 Simulations show that our framework outperforms the popular technique of \gls{mse} minimization~\cite{stoica_probing_2007,liu_mu-mimo_2018,liu_toward_2018,liu_joint_2020}.

In the remainder of this section, we introduce the notation and our system model. We formalize the optimization problem for analog-beamforming \gls{isac} detection in \cref{sec:problem}, for which we derive parallel \gls{tx}/\gls{rx} problems that we approximate via superiorized projections onto convex sets.
We deliver numerical results in \cref{sec:results}, and a conclusion in \cref{sec:conclusions}.

\subsection{Notation and Preliminaries}

In the following, lower case letters $a$ denote scalars, bold lower case letters $\x$ denote column vectors, and bold upper case letters $\X$ denote matrices. We write the complex conjugate of a scalar $z\in\CC$ as
$z^*$, the transpose of a vector or matrix $\X$ as $\X^T$, and its Hermitian transpose as $\X^H$. We denote by $\I$, $\Null$ and $\e_m$ the identity matrix, zero matrix, and the $m$th Cartesian unit vector, respectively, where the dimensions will be clear from the context.
We write ($\forall\x\in\CC^N$) $\|\x\|\coloneqq\sqrt{\x^H\x}$ for the Euclidean norm of $\x$,
and $\EE[X]$ for the expected value of a random variable $X$.
We also write $\Rp\coloneqq(0,\infty)$ for the set of the positive real numbers, 
$\delta\coloneqq\ZZ\rightarrow\lbrace0,1\rbrace$ for the Kronecker delta with $\delta(n)=1$ if $n=0$ and $\delta(n)=0$ otherwise,
and $\X\succeq\Null$ for
positive semidefinite matrices $\X$.
We use the shortcuts
$[N]\coloneqq\lbrace1,\ldots,N\rbrace$ for positive index sets with $N\in\NN$,
and $\VV\subseteq[-\pi,\pi)$ for the visible region of an antenna array.

For any closed convex set $\setC$ in a Hilbert space $(\hilbert,\langle\cdot,\cdot\rangle)$, we denote by $P_{\setC}(\X)$ the projection of $\X\in\hilbert$ onto $\setC$, and by
$(\forall\X\in\hilbert)$ $T_\setC^\mu(\X)\coloneqq
\X+\mu(P_\setC(\X)-\X)$ the relaxed projection onto $\setC$
with a relaxation parameter $\mu\in(0,2)$.
Additionally, we use the shortcut ($\forall l\in[L],\,\forall\setC_l\subset\hilbert$) $T_{\setC_{L\leftarrow1}}^\mu(\X)\coloneqq
T_{\setC_L}^\mu\ldots T_{\setC_1}^\mu(\X)$. Throughout this paper, we consider
the {\emph{real}} Hilbert space of Hermitian matrices,
$\hilbert=\HH^M\coloneqq\lbrace\X\in\CC^{M\times M}|
\,\X=\X^H\rbrace$ with the trace inner product $\langle\A,\B\rangle\coloneqq\trace(\B^H\A)={\trace(\B\A)\in\RR}$,
which induces the Frobenius norm
$(\forall\X\in\hilbert)$ $\|\X\|_\hilbert\coloneqq\sqrt{\langle\X,\X\rangle}$.

\subsection{System Model}\label{sec:sysmodel}

\begin{figure*}[!t] 
    \centering
      \includegraphics[width=\linewidth]{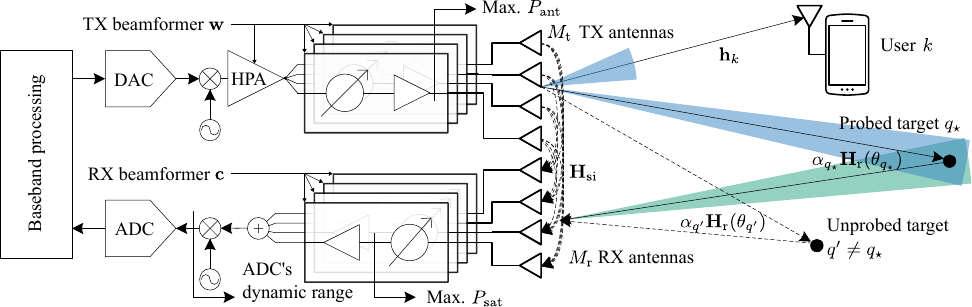}
    \caption{System model of a monostatic \gls{isac} transceiver
    with analog beamforming.}
    \label{fig:sysmodel}
\end{figure*}

The system model is depicted in
\cref{fig:sysmodel}. It consists of an \gls{isac} transceiver with
$\Mt$ and $\Mr$ antennas for \gls{tx} and \gls{rx} and individual
phase and gain control.
The \gls{tx} beamformer
$\w\in\CC^{\Mt\times1}$ is not normalized to account for the 
\gls{tx} power $\Pt\coloneqq\|\w\|^2$,
while the \gls{rx} beamformer
$\c\in\CC^{\Mr\times1}$ fulfills $\|\c\|=1$.
Each \gls{tx} antenna has a maximum output power, $\Pant$, and
each \gls{rx} antenna has a saturation power, $\Psat$.

At a given time slot of duration $T$, the \gls{isac} transceiver
produces a narrowband baseband signal
$(\forall n\in[N])$ $ s(n)\in\CC$ with i.i.d. samples and
unit average power, e.g.:
\begin{equation}
(\forall n\in[N])\qquad
\EE\left[|s(n)|^2\right]=1\label{eq:signal}\,.
\end{equation}
Under~\eqref{eq:signal}, the antenna covariances can be computed
as $\W\coloneqq\w\w^H$ and $\C\coloneqq\c\c^H$~\cite{liu_mu-mimo_2018}.
The signal is transmitted over a
carrier frequency $f_0$ to serve multiplexed data to $K$ users.
For this, a bandwidth $B_k$ is allocated to each user so that the total bandwidth
$B_s\coloneqq\sum_{k\in[K]}B_k<N/T\ll f_0$.
After reception and demodulation, each user obtains
$\left(\forall k\in[K]\right)$
${y}_k(n)=\bar{y}_k(n) + w_k(n)$,
where $\bar{y}_k(n)\coloneqq\h_k^H\w s(n)$, $\h_k\in\CC^{\Mt\times1}$
is the MIMO channel for user $k$, and
$\left(\forall n\in[N]\right)$
$w_k(n)\sim\CN(0,\sigma_k^2)$ circularly-symmetric complex white Gaussian
noise. Each user $k$ requires a certain rate $R_k$ that is achieved by meeting a minimum \gls{snr} $\Gamma_k$ $(\forall k\in[K])$:
\begin{equation}
    \frac{\EE[|\bar{y}_k(n)|^2]}{\EE[|w_k(n)|^2]}=
    \frac{|\h_k^H\w|^2}{\sigma_k^2}\ge\Gamma_k=2^{R_k/B_k}-1\,.\label{eq:snr}
\end{equation}

\subsubsection*{Integrated Sensing and Communication}
The \gls{isac} transceiver also receives $s(n)$ via the backscatter and \gls{si}
channels with the impulse responses given by ($\forall n\in[N]$) $\bHr(n),\bHsi(n)\in\CC^{\Mr\times\Mt}$,
respectively.
$\bHr(n)$ originates from $Q$ targets, each one
characterized by the tuple $(\xi_q,r_q,\theta_q)\in
\Rp\times\Rp\times\VV$ representing
the radar cross section, the distance to the transceiver, and the direction of arrival:
\begin{subequations}
\begin{align}
    \bHr(n)&=
    \sum_{q\in[Q]}{
    \alpha_q
    \Hr(\theta_q)\delta\left(n-n_q\right)}\,,\;
    n_q\coloneqq\frac{2Nr_q}{Tc},\\
    \Hr(\theta_q)&=\aR(\theta_q)\aT^H(\theta_q)\,,
    \label{eq:radch}\qquad\;
    \alpha_q=\sqrt{\frac{\xi_q\lambda^2}{(4\pi)^3r_q^4}}\,.
\end{align}
\end{subequations}
Here and hereafter, $c$ is the speed of light, $\lambda=c/f_0$ the wavelength, and
($\forall\theta\in\VV$)
$\aT(\theta)\in\CC^\Mt,\aR(\theta)\in\CC^\Mr$ are
the \gls{tx} and \gls{rx} steering vectors.
We assume enough proximity between \gls{tx} and \gls{rx}
antennas to consider $\bHsi(n)$ an instantaneous flat-fading channel, $\bHsi(n)\approx\Hsi\delta(n)$,
$\Hsi\in\CC^{\Mr\times\Mt}$,
and hence we can write the signal received by the
transceiver as ($\forall n\in[N]$)
\begin{align}\label{eq:yr}
    \yr(n)&=\bar{y}_\mathrm{r}(n)+\wr(n),\quad\wr(n)\sim\CN(0,\sigr^2),\\
    \bar{y}_\mathrm{r}(n)&=\sum_{q\in[Q]} \nonumber
    \alpha_q
    \c^H\Hr(\theta_q)\w s\left(n-n_q\right)+\c^H\Hsi\w s(n)\;.
\end{align}

Without loss of generality, we assume $M\coloneqq\Mt=\Mr$ and $\a(\theta)\coloneqq\aT(\theta)=\aR(\theta)$
to ease the formulation in the remainder of the paper. We also note that the model can be easily extended to incorporate the targets' velocities $\nu_q$ by redefining the one-dimensional $\delta(n-n_q)$ into a 2D delta in the
delay-Doppler plane, $\delta'(n-n_q,\,\nu-\nu_q)$
\cite{dehkordi_beam-space_2023}.
 Likewise, bidirectional communication can be accommodated through time division,
 for which the transceiver's \gls{rx} beamformer can be optimized in a completely decoupled way.

%% file: content/problem.tex
The objective of the proposed algorithm is the design of the analog beamforming vectors $\w$ and $\c$ that optimize (in a well-defined sense) the \gls{isac}
operation of the transceiver in \cref{fig:sysmodel}.
To this end, we must take into account the power limitations $\Pant$, $\Psat$ from \cref{sec:sysmodel},
 and also impose \gls{si}-suppression capabilities, $\c^H\Hsi\w\simeq0$.
Moreover, we need to incorporate the inequalities in~\labelcref{eq:snr} to ensure the desired communication performance. However, we first need to define a notion of sensing performance for which $\w$ and $\c$ can be optimized.

\subsection{Analog Beamforming and Radar Sensing}
\label{sec:detect}

As mentioned in~\cref{sec:intro}, we address sensing through target detection rather than angle estimation. In the literature on
\gls{isac} beamforming, angle estimation is often approached by minimizing the \gls{mse} of $\W$
either with respect to a desired covariance $\W_*$~\cite{liu_mu-mimo_2018,liu_toward_2018},
 or to a beampattern $p(\theta)$, $p:\,\VV\rightarrow\Rp\cup\lbrace 0\rbrace$~\cite{stoica_probing_2007, liu_joint_2020}, i.e.:
\begin{equation}
    L_{\textrm E}(\W,\zeta)\coloneqq\frac{1}{|\THe|}
    \sum_{\theta\in\THe}\left|\zeta p(\theta)-
        \a(\theta)^H\W\a(\theta)\right|^2\,.\label{eq:mse}
\end{equation}
Here, $\zeta\in\Rp$ is an auxiliary optimization variable
and $\THe\subset\VV$ is a finite set of angle grid points.
Alternatively, \citet{liu_cramer-rao_2022} directly optimize
the Cram\'{e}r-Rao bound for angle estimation under digital \gls{tx} beamforming.

On the other hand, array-processing techniques are not available for analog beamforming,
and the angle must be typically estimated through
beam sweeping, possibly combined with a
discovery/tracking scheme~\cite{dehkordi_beam-space_2023}.
In this setup, the beam's main-lobe width $\Delta\theta$ is the limiting factor for
angle estimation, which is in turn mostly restricted by $M$.

Target detection offers more design freedom for $\w$ and $\c$. In our multi-target scenario, we define an angular bin
$\Theta_b\coloneqq[\theta_b-\Delta\theta/2, \theta_b+\Delta\theta/2]\subset\VV$, characterized by a central angle $\theta_b$ and beam width $\Delta\theta$, to probe the presence of a target $\qstar\in[Q]$. Noting that
$\lbrace\theta_q\rbrace^Q_ {q=1}\cap\Theta_b=\lbrace\theta_\qstar\rbrace$ and assuming
\gls{si} cancellation in \labelcref{eq:yr}, we can pose the following hypothesis test:
\begin{align}
\begin{split}
    \Hnull: \;     \yr(n)&=
    \sum_{\substack{q\in[Q],\\\theta_q\notin\Theta_b}}{
    \alpha_q
    \c^H\Hr(\theta_q)\w s\left(n-n_q\right)}
   +\wr(n),\\
    \Halt:  \; \yr(n)&=
    \sum_{q\in[Q]}{
    \alpha_q
    \c^H\Hr(\theta_q)\w s\left(n-n_q\right)}
   +\wr(n)\,,\label{eq:ht-mtgt}
\end{split}
\end{align}
with unknown $Q$, $\alpha_q$, $n_q$, and $\theta_q$.
A powerful test statistic for \labelcref{eq:ht-mtgt} is elusive due to the lack of knowledge on most parameters~\cite{kay_fundamentals_2009}. Luckily, we can design
$\w$ and $\c$ such that
\begin{equation}
    (\forall\theta\in\Theta_b)(\forall\vartheta\in\VV\setminus\Theta_b)\;
    |\c^H\Hr(\theta)\w|\gg|\c^H\Hr(\vartheta)\w|\,.\label{eq:ht-sll}
\end{equation}
In that case, we can approximate \labelcref{eq:ht-mtgt} as
\begin{align}
\begin{split}
    \Hnull: \; \yr(n)&=\wr(n)\;,\\
    \Halt:   \;\yr(n)&=\alpha_\qstar\c^H\Hr(\theta_\qstar)\w s(n-n_\qstar) + \wr(n)
    \;.
\end{split}
\end{align}

We can now correlate $\yr(n)$ with the known signal $s(n)$
and apply the \acrlong{glrt}~\cite{kay_fundamentals_2009} ($\forall n\in\ZZ\setminus[N]$ $s(n)=0$):
\begin{equation}\label{eq:corr-ht}
    \frac{\max_{n'\in[N]}\left|\sum_{n\in[N]}{\yr(n)s^*(n-n')}\right|^2}
    {N\sigr^2/2}
    \underset{\Hnull}{\overset{\Halt}{\gtrless}}
    \eta\,,
\end{equation}
where $\eta\in\Rp$ is a parameter that trades off the probability of false alarm $\Pfa$ and the probability of detection $\Pd$. Specifically, $\Pfa$ and $\Pd$ are given by $\chisq$ and $\chisq(\rho)$, 
the central and non-central chi distributions with two degrees of freedom, respectively, and
$\rho$ the non-centrality parameter
\begin{equation}
\rho=\frac{\alpha_\qstar^2N|\c^H\Hr(\theta_\qstar)\w|^2}{\sigr^2/2}\,.\label{eq:radsnr}
\end{equation}
It is known that, for a fixed $\Pfa$,
$\Pd\eqqcolon f_\mathrm{D}\left(\rho\;;\Pfa\right)$
is monotonically increasing in $\rho$~\cite{kay_fundamentals_2009}.
In other words, we can design
$\w$ and $\c$ to maximize $\Pd$ via
\labelcref{eq:radsnr} while enforcing
\labelcref{eq:ht-sll}, which leads to a low-\gls{sll} design as in~\cite{liu_full-duplex_2023}. 

\subsection{Optimization Problem Statement}

Taking into account the previous discussion,
we pose the following optimization problem:
\begin{subequations}\label{eq:prob}
\begin{align}
    \underset{\w\in\CC^M,\c\in\CC^M}{\maximize}\,\min_{\theta\in\THl}&|\c^H\Hr(\theta)\w|
    \label{eq:mmlobe}\qquad\st\\
    \left(\forall(\theta,\vartheta)\in\THl\times\THs\right)\;&{{|\c^H\Hr(\theta)\w|}}\ge\gams{|\c^H\Hr(\vartheta)\w|}
    \label{eq:sll}\\
    &\c^H\Hsi\w=0\label{eq:si}\\
    (\forall k\in[K])\quad&
    \frac{|\h_k^H\w|^2}{\sigma_k^2}\geq\Gamma_k
    \label{eq:snr-opt}\\
    (\forall m\in[M])\quad
    &\e_m^H\w\w^H\e_m\le\Pant\label{eq:pant}\\
    (\forall m\in [M])\quad&
    \gm^H\w\w^H\gm\le\Psat
    \label{eq:psat}\\&
    \|\c\|=1\;,
\end{align}
\end{subequations}
where $\THl\coloneqq\lbrace\theta_1,\ldots,\theta_I\rbrace\subset\Theta_b$ and
$\THs\coloneqq\lbrace\vartheta_1,\ldots,\vartheta_J\rbrace\subset\VV\setminus\Theta_b$
are two finite sets of angle points
in the main lobe and sidelobes, respectively,
$\gm\coloneqq\Hsi^H\e_{m}$,
and $\gams$ is a target
\gls{sll}.
We note that  \labelcref{eq:sll} enforces \labelcref{eq:ht-sll} under proper $\gams$, $\THl$ and
$\THs$, in which case
\labelcref{eq:mmlobe} approximates optimal multi-target detection via \labelcref{eq:radsnr}.
Moreover, the \gls{si}-suppression and \gls{snr} requirements are imposed by \labelcref{eq:si} and \labelcref{eq:snr-opt},
while \labelcref{eq:pant,eq:psat} impose the antenna power limitations due to $\Pant$ and $\Psat$.

The joint problem \labelcref{eq:prob} is a nonconvex quadratically constrained quadratic program and thus NP-hard~\cite{fink_multi-group_2021}.
In particular, the coupling of
$\w$ and $\c$ in \labelcref{eq:mmlobe,eq:sll,eq:si}
renders the optimization problem difficult to solve.
We suggest thus splitting \labelcref{eq:prob} into
two parallel optimization problems for
\gls{tx} and \gls{rx}. For this, we first use
the singular value decomposition of
$\Hsi=\U\S\V^H$ to write:
    \begin{align}
    \Hsi&=\Hu\Hv,\quad
    \Hu\coloneqq\U\S^{1/2},\quad
    \Hv\coloneqq\S^{1/2}\V^H\,,\\
    \S^{1/2}&\coloneqq\diag(\sqrt{[\S]}_{1,1},\,
\ldots\,,\sqrt{[\S]}_{M,M})\;.\nonumber
\end{align}
Additionally,
we recall \labelcref{eq:radch} to note that
\begin{align}
    ( \forall\theta\in\VV)\quad
    |\c^H\Hr(\theta)\w|&=|\c^H\a(\theta)||\a^H(\theta)\w|\,,
   \label{eq:jnt-bf}
\end{align}
and hence we write the two separate problems \labelcref{eq:prob-tx,eq:prob-rx}:

\begin{subequations}\label{eq:prob-tx}
\begin{align}
    \underset{\w\in\CC^M}{\maximize}\min_{\theta\in\THl}&|\a^H(\theta)\w|\label{eq:mmlobe-tx}\qquad\st\\
    (\forall(\theta,\vartheta)\in\THl\times\THs)\quad&
    {|\a^H(\theta)\w|}\ge\sqrt{\gams}{|\a^H(\vartheta)\w|}\label{eq:sll-tx}\\&
       \|\Hv\w\|=0\label{eq:si-tx}\\
       &\text{\crefrange{eq:snr-opt}{eq:psat}}
\end{align} and
\end{subequations}
\vspace*{-5mm}
\begin{subequations}\label{eq:prob-rx}
\begin{align}
    \underset{\c\in\CC^M}{\maximize}\min_{\theta\in\THl}&|\c^H\a(\theta)|\label{eq:mmlobe-rx}\qquad\st\\
   (\forall(\theta,\vartheta)\in\THl\times\THs)\quad&{|\c^H\a(\theta)|}\ge\sqrt{\gams}{|\c^H\a(\vartheta)|}\label{eq:sll-rx}\\
    &\|\c^H\Hu\|=0\label{eq:si-rx}\\
    &\|\c\|=1\label{eq:mod-rx}\;.
\end{align}
\end{subequations}

Let us write $\Fj\subset\CC^{M}\times\CC^{M}$ for the feasible region of \labelcref{eq:prob}, and
$(\wstar,\cstar)$ for a solution of \labelcref{eq:prob-tx}--\labelcref{eq:prob-rx}.
It is clear that \labelcref{eq:si-tx} \& \labelcref{eq:si-rx} imply \labelcref{eq:si}, and
\labelcref{eq:sll-tx} \& \labelcref{eq:sll-rx} imply \labelcref{eq:sll},
so that $(\wstar,\cstar)\in\Ftr\subset\Fj$.
It can also be proven that $(\wstar,\cstar)$,
is a solution of \labelcref{eq:mmlobe} in $\Ftr$ if the following condition holds:
\begin{equation}\label{eq:check}
    \argmin_{\theta\in\THl}{|\a^H(\theta)\wstar|}
    {=}
    \argmin_{\theta\in\THl}{|\cstar^H\a(\theta)|}.
\end{equation}

\subsection{Superiorized Projections in Hilbert Spaces}

While \labelcref{eq:prob-tx,eq:prob-rx}
are still nonconvex, we can use the trace identity $\trace(\A\B)=\trace(\B\A)$ and its particularization
$\trace(\v\v^H)=\v^H\v$ to obtain semidefinite reformulations~\cite{fink_multi-group_2021,liu_mu-mimo_2018,liu_toward_2018,liu_joint_2020}.
That is, we define ($\forall\theta\in\VV$) $\A_\theta\coloneqq\a(\theta)\a^H(\theta)$, square \labelcref{eq:mmlobe-tx,eq:sll-tx,eq:si-tx,eq:mmlobe-rx,eq:sll-rx,eq:si-rx,eq:mod-rx}, rearrange, and rewrite
\labelcref{eq:prob-tx,eq:prob-rx} as
\begin{subequations}\label{eq:sdr-tx}
\begin{align}
    \underset{\W\in\HH^M}{\maximize}\min_{\theta\in\THl}\;&
    \trace(\A_\theta\W)\label{eq:sdt-mmlobe}\qquad\st\\
    (\forall(\theta,\vartheta)\in\THl\times\THs)\quad&
    \trace\left((\A_\theta-{\gams}\A_\vartheta)\W\right)\geq0\label{eq:sdt-sll}\\
    &\trace(\Hv^H\Hv\W)=0\label{eq:sdt-si}\\
    (\forall k\in[K])\quad&
    \trace(\h_k\h_k^H\W)\geq\sigma_k^2\Gamma_k\label{eq:sdt-snr}\\
    (\forall m\in[M])\quad&
    \trace(\e_m\e_m^H\W)\le\Pant\label{eq:sdt-pant}\\
    (\forall m\in[M])\quad&
    \trace(\gm\gm^H\W)\le\Psat\label{eq:sdt-psat}\\
    &\W\succeq\Null\label{eq:sdt-psd}\\
    &\rank(\W)\leq1\label{eq:ranktx}
\end{align}
\end{subequations}
\vspace*{-5mm}
\begin{subequations}\label{eq:sdr-rx}
\begin{align}
    \underset{\C\in\HH^M}{\maximize}\min_{\theta\in\THl}\;&\trace(\A_\theta\C)\label{eq:sdr-mmlobe}\qquad\st\\
    (\forall(\theta,\vartheta)\in\THl\times\THs)\quad&
    \trace\left((\A_\theta-{\gams}\A_\vartheta)\C\right)\geq0\label{eq:sdr-sll}\\
    &\trace(\Hu\Hu^H\C)=0\label{eq:sdr-si}\\
    &\trace(\C)=1\label{eq:sdr-mod}\\
    &\C\succeq\Null\label{eq:sdr-psd}\\
    &\rank(\C)\leq1\;.\label{eq:rankrx}
\end{align}
\end{subequations}
The optimization constraints $\W,\C$ belong now to the Hilbert space $\hilbert$, and the objective functions are minima of linear expressions, so that 
\labelcref{eq:sdr-mmlobe,eq:sdt-mmlobe} maximize concave functions.
Moreover, \labelcref{eq:ranktx,eq:rankrx} represent the closed set
$\setR\coloneqq\lbrace
    \X\in\hilbert\,|\,\rank(\X)\leq1\rbrace$, the only nonconvex constraint,
while \labelcref{eq:sdt-sll,eq:sdt-si,eq:sdt-snr,eq:sdt-pant,eq:sdt-psat,eq:sdt-psd,eq:sdr-sll,eq:sdr-si,eq:sdr-psd,eq:sdr-mod} represent closed convex sets in $\hilbert$
($\forall(\theta,\vartheta)\in\THl\times\THs$)
($\forall k\in[K]$) ($\forall m\in[M]$):
\begin{subequations}
\begin{align}
    \setV&\coloneqq\lbrace\label{eq:setV}
    \X\in\hilbert\,|\,\langle\Hv^H\Hv,\X\rangle= 0\rbrace\,,\\
     \setN_{k}&\coloneqq\lbrace\label{eq:setN}
    \X\in\hilbert\,|\,\langle\h_k\h_k^H,\X\rangle\geq \sigma_k^2\Gamma_k\rbrace,\\
    \setA_{m}&\coloneqq\lbrace\label{eq:setA}
    \X\in\hilbert\,|\,\langle\e_m\e_m^H,\X\rangle\leq \Pant\rbrace\,,\\
    \setS_{m}&\coloneqq\lbrace\label{eq:setS}
    \X\in\hilbert\,|\,\langle\gm\gm^H,\X\rangle\leq \Psat\rbrace\,,\\
    \setL_{(\theta,\vartheta)}&\coloneqq\lbrace\label{eq:setL}
    \X\in\hilbert\,|\,\langle\A_\theta-{\gams}\A_\vartheta,\X\rangle\geq 0\rbrace,\\
    \setP&\coloneqq\lbrace\label{eq:setP}
    \X\in\hilbert\,|\,\X\succeq\Null\rbrace\,,\\
    \setU&\coloneqq\lbrace\label{eq:setU}
    \X\in\hilbert\,|\,\langle\Hu\Hu^H,\X\rangle= 0\rbrace\,,\\
    \setI&\coloneqq\lbrace\label{eq:setI}
    \X\in\hilbert\,|\,\langle\I,\X\rangle= 1\rbrace\,.
\end{align}
\end{subequations}
In order to find solutions to \labelcref{eq:sdr-tx,eq:sdr-rx},
we first drop the objective functions
and rank constraints
to consider the following convex feasibility problems:
\begin{align}
    \text{Find }&\W\in\hilbert\text{ such that $\W$ in
    \labelcref{eq:setV,eq:setN,eq:setA,eq:setS,eq:setL,eq:setP},}
    \label{eq:feas-tx}\\
    \text{Find }&\C\in\hilbert\text{ such that $\C$ in \labelcref{eq:setU,eq:setI,eq:setL,eq:setP}.}
    \label{eq:feas-rx}
\end{align}
Then, we approximate a solution to \labelcref{eq:feas-tx,eq:feas-rx} through iterative
sequential projections, where we only use two distinct relaxation parameters $\mu,\mu'$ for simplicity:
\begin{align}
    (\forall i\in\NN)\;\;
    &\W_{i+1}=T_\mathrm{T}(\W_i)\label{eq:projW}\\
    &\coloneqq
    T_{\setP}^{\mu'}
    T_\setV^{\mu'}
    T_{\setN_{K\leftarrow1}}^\mu
    T_{\setL_{\substack{\theta_I\leftarrow\theta_1\\
            \vartheta_J\leftarrow\vartheta_1}}}^\mu
    T_{\setS_{M\leftarrow1}}^\mu
    T_{\setA_{M\leftarrow1}}^\mu
    (\W_i)\nonumber\\
    (\forall i\in\NN)\;\;&
    \C_{i+1}=T_\mathrm{R}(\C_i)\label{eq:projC}\\
    &\coloneqq
    T_{\setP}^{\mu'}
    T_\setU^{\mu'}
    T_\setI^{\mu'}
    T_{\setL_{(\theta_I,\vartheta_J)}}^\mu\ldots
    T_{\setL_{(\theta_1,\vartheta_1)}}^\mu
    (\C_i)\,.\nonumber
\end{align}
The mathematical expressions of the projections can be found in~\cite{fink_multi-group_2021,theodoridis_adaptive_2011}.
It has been shown in \cite{fink_multi-group_2021,he2017perturbation} that the mappings $T_\mathrm{T}$ and $T_\mathrm{R}$ are \emph{bounded perturbation resilient}, i.e., they generate sequences converging to solutions of the convex feasibility problems in \labelcref{eq:feas-rx,eq:feas-tx}, even if bounded perturbations are added to their iterates. This enables us to use the iterations in \labelcref{eq:projW,eq:projC} as \emph{basic algorithms} for superiorization. 
By adding a sequence of well-designed bounded perturbations\footnote{See \cite{censor2014weak} for a formal definition of bounded perturbations.} $(\beta_i\v_i)_{i\in\NN}$ to the iterates of a basic algorithm $\x_{i+1} = T(\x_i)$, $\x_i\in\setH$, the superiorization methodology \cite{censor2010perturbation} automatically produces its \emph{superiorized version} $\x_{i+1} = T(\x_i+\beta_i\v_i)$, $\x_i\in\setH$, which aims to find feasible points that are superior in a specified sense. 
Hence, the sequences produced by superiorized versions of the iterations in \labelcref{eq:projW,eq:projC} are guaranteed to converge to solutions to \labelcref{eq:feas-rx,eq:feas-tx}, respectively.

In the following, we devise a sequence of perturbations to increase the objective value in \eqref{eq:sdt-mmlobe} and \eqref{eq:sdr-mmlobe}, while simultaneously reducing the distance to the rank-constraint set $\setR$.
For this, we borrow from~\cite[Eq. (15)]{fink_multicast_2019} the perturbation
$\setK(\X)\coloneqq P_\setR(\X)-\X$,
where $P_\setR(\X)$ is a 
projection onto $\setR$, and
we conceive the perturbation $\setY(\X)$ for the supergradients of the objective functions:
\begin{align}
    \setY(\X)&\coloneqq\left(v_{\bar{\theta}}-v_{\underline{\theta}}\right)\frac{\A_{\underline{\theta}}}{\|\A_{\underline{\theta}}\|_\hilbert}\,,\quad\label{eq:mmin-pert}
    v_{{\theta}}\coloneqq\frac{\langle\A_\theta,\X\rangle}{\|\A_{{\theta}}\|_\hilbert},\\
    \bar{\theta}&\in\argmax_{\theta\in\THl}{v_\theta}\;,\quad
    \underline{\theta}\in\argmin_{\theta\in\THl}{v_\theta}\;.\nonumber
\end{align}
In words, $\setY(\W)$ and $\setY(\C)$ increase the smallest term inside \labelcref{eq:sdt-mmlobe,eq:sdr-mmlobe}, respectively, up to the largest one.
Since \labelcref{eq:sdt-mmlobe,eq:sdr-mmlobe} are proportional to $\|\W\|_\hilbert$ and $\|\C\|_\hilbert$, we should ensure that
$\W$ and $\C$ reach their maximum feasible norms.
Noting that \labelcref{eq:sdr-mod} already fixes $\|\C\|_\hilbert$,
we add a scale perturbation just for $\W$:
\begin{align}\label{eq:scale}
    \setB(\W)&\coloneqq\left(\beta-1\right)\W\,,\\
    \beta&\coloneqq\min\left(\min_{m\in[M]}{\frac{\Pant}{\langle\e_m\e_m^H,\W\rangle}},\min_{m\in[M]}{\frac{\Psat}{\langle\gm\gm^H,\W\rangle}}\right).
    \nonumber
\end{align}

The final algorithm iterations are as follows:
\begin{align}
    \W_{i+1}&=T_\mathrm{T}(\W_i+
    c_1^i\setK(\W_i)+c_2^i\setY(\W_i)+c_3^i\setB(\W_i))\nonumber\\
    \C_{i+1}&=T_\mathrm{R}(\C_i+
    c_1^i\setK(\C_i)+c_2^i\setY(\C_i))\,,\label{eq:algo}
\end{align}
with $a_1, a_2, a_3\in(0,1)$ the decaying coefficients of the perturbations.
Since \labelcref{eq:algo} is guaranteed to converge,
we set the stop condition
$\|\X_{i+1}-\X_i\|_\hilbert<\epsilon\|\X_{i+1}\|_\hilbert$
and extract $(\wstar,\cstar)$ as the largest principal component from $\W_i$ and $\C_i$.

%% file: content/results.tex
For the simulations, we choose $f_0=\text{\SI{28}{\giga\hertz}}$
and $\lambda/2$-spaced \glspl{ula} with
$\Pant=\text{\SI{10}{\dBm}}$, 
$\Psat=\text{\SI{-20}{\dBm}}$, 
and $\Mt=\Mr=10$, so that $\VV=[-\pi/2,\pi/2]$ and
\begin{equation}
    \a(\theta)=\left[1,e^{j\pi\sin{\theta}},\ldots,e^{j\pi(M-1)\sin{\theta}}\right]^T,\quad
    M=10\,.
\end{equation}
Moreover, we assume that the \gls{tx} and \gls{rx}
arrays constitute the left and right subarrays of
a $\lambda/2$-\gls{ula} of size $\Mt+\Mr$ and
total length \SI{10.2}{\centi\meter}, for which we model $\Hsi$ as:
\begin{equation}
    [\Hsi]_{m',m}=\varrho(d_{m',m})\exp{(2\pi d_{m',m}/\lambda)}\,,
\end{equation}
where $d_{m',m}$ is the distance between \gls{tx} and \gls{rx} antennas $m$ and $m'$, and $\varrho(d)$ is Friis' path loss, which is a valid approximation for $d_{m',m}\geq\lambda/2$~\cite{schantz_near_2005}.

We consider ($\forall q\in[Q]$)
$\xi_q=\text{\SI{1}{\meter\squared}}$,
($\forall k\in[K]$)
$\sigma_k^2=\sigr^2=\text{\SI{-88}{\dBm}}$, and
$\Gamma_k=\text{\SI{3}{\deci\bel}}$.
We adopt the geometric fading channel model from~\cite{liu_joint_2023} for $\h_k$ with
path-loss factor 2.2
and Rician factor $\kappa=1000$.

\subsection{Beamforming Analysis}

We choose a beam at
$\theta_b=\text{\SI{0}{\degree}}$ with
$\Delta\theta=\text{\SI{20}{\degree}}$,
which gives $\gams\simeq\text{\SI{30}{\deci\bel}}$ according to
Dolph-Chebyshev's minimum \gls{sll} for $M=10$.~\cite{lynch_dolphchebyshev_1997}.
We also leave a \SI{6}{\degree} guard band, so that
$\THl$ spans \SIrange{-7}{7}{\degree},
and $\THs$ spans \SIrange{\pm13}{\pm90}{\degree}, in \SI{1}{\degree}-steps.
For the \gls{spocs}, we set $\mu'=1$, $\mu=1.5$, $\epsilon=10^{-5}$,
$a_1=a_3=0.9999$, and $a_2=0.99$.
We obtain $\wstar$ for $K=2$ after 1000 iterations, where the users are located
at $\theta_1=\text{\SI{-25}{\degree}}$, $\theta_2=\text{\SI{40}{\degree}}$, and $r_k=\text{\SI{20}{\meter}}$ for $k=1,2$.

\begin{figure}[!t]
     \centering
     \begin{subfigure}
         \centering
         \includegraphics[width=\linewidth]{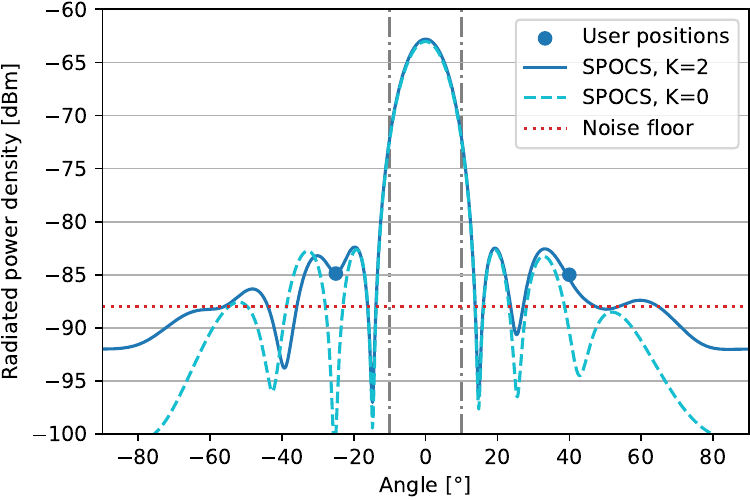}
         \caption{\gls{tx} \acrlong{aps} at $r_k=\text{\SI{2}{\meter}}$. The inclusion of $K=2$ users alters the sidelobes to guarantee their \gls{snr}.}
         \label{fig:tx-bf}
     \end{subfigure}
     \begin{subfigure}
         \centering
         \includegraphics[width=\linewidth]{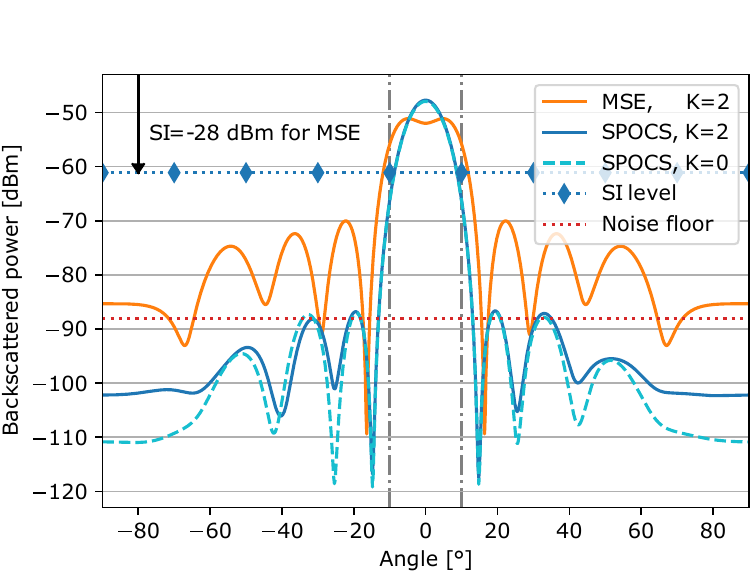}
         \caption{Backscattered \acrshort{aps} from target $\qstar$ at $r_\qstar=\text{\SI{2}{\meter}}$. Our method decreases \gls{sll} by \SI{20}{\deci\bel} and adds \SI{33}{\deci\bel} \gls{si} attenuation compared to \gls{mse}.}
         \label{fig:joint}
     \end{subfigure}
\end{figure}

\cref{fig:tx-bf} shows the
\gls{aps} at $r_k$ for $K=2$ and compares it with
a userless case ($K=0$). The main beam and \gls{sll}
are similar in both cases, but the \gls{snr} constraints in \labelcref{eq:snr} shift and fill the beampattern's zeros for $K=2$.
In practice, $\gams$ needs to be selected carefully to avoid conflicts with the required \glspl{snr} $\Gamma_k$.

The overall effect of $(\wstar,\cstar)$ is shown in \cref{fig:joint} for a target
$\qstar$ at $r_\qstar=\text{\SI{2}{\meter}}$ and arbitrary $\theta_\qstar$.
The joint beamforming fulfills \labelcref{eq:check} and attains an \gls{si} suppression below
\SI{-60}{\dBm}, which leaves only \SI{27}{\deci\bel} above the noise floor to be canceled digitally.
We compare its \gls{aps} with the baseline from~\cite{liu_joint_2020}. For this,
 we set \gls{mse} minimization from
\labelcref{eq:mse} as the objective in \labelcref{eq:sdr-tx,eq:sdr-rx}
with $\THe=\THl\cup\THs$ and $p(\theta)=1$ for $\theta\in\Theta_b$ and $0$ elsewhere;
we enforce \labelcref{eq:sdt-pant} with equality,
we drop all \gls{si} and \gls{sll} constraints, 
and we obtain $\W$ and $\P$ with a standard convex solver.
The \gls{mse} baseline yields an \gls{si} level of \SI{-28}{\dBm},
\SI{33}{\deci\bel} above our method, which represents a reduction in the \gls{adc}'s dynamic range equivalent to 5 bits.
Moreover, the baseline's \gls{sll} lies \SI{20}{\deci\bel} above that of
\gls{spocs}.

\subsection{Multi-target Detection}

We investigate the impact of \gls{sll} on detection
in a scenario with one probed and one unprobed target, $\qstar$ and $q'$, with
$r_\qstar=\text{\SI{20}{\meter}}$,
$r_{q'}\in\lbrace2,5,10\rbrace$\si{\meter},
$\theta_\qstar\in\Theta_b$,
and $\theta_{q'}\in\VV\setminus\Theta_b$.
We estimate $\Pd$ and $\Pfa$
for $(\wstar,\cstar)$ from \labelcref{eq:corr-ht},
where we use 100
log-spaced values for
$\eta\in[5,\,5\times10^6]$, we run 5000 Monte Carlo simulations per $\eta$ value,
and we sample $\theta_{\qstar}$ and $\theta_{q'}$ uniformly.
The waveform of $s(n)$ is single-carrier \gls{ofdm} with
\acrshort{qpsk} modulation,
5G numerology 5, and 68
resource blocks; hence
$T=\text{\SI{2.23}{\micro\s}}$, $N=4384$,
and $B_s=\text{\SI{391.7}{\mega\hertz}}$.

 \begin{figure}[t]
         \centering
     \includegraphics[width=\linewidth]{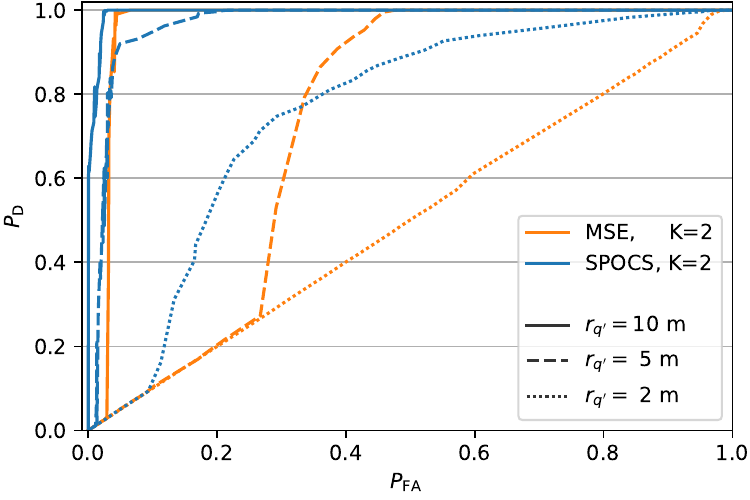}
\caption{\acrshort{roc} curves with $r_\qstar=\text{\SI{20}{\meter}}$ and unprobed target $q'$.}
\label{fig:roc}
 \end{figure}

According to the resulting \gls{roc} curves in \cref{fig:roc},
$r_{q'}=\text{\SI{10}{\meter}}$ has little impact on detection for both beampatterns,
but \gls{mse}'s performance sharply decreases for closer unprobed targets.
Indeed, \gls{spocs} achieves perfect $\Pd$ at $r_{q'}=\text{\SI{5}{\meter}}$
for $\Pfa>0.2$ in contrast to $\Pfa>0.45$ for \gls{mse}.
Furthermore, $q'$ completely overshadows the \gls{mse}
beampattern at \SI{2}{\meter}, so that $\Pd\simeq\Pfa$ $\forall\eta$, whereas \gls{spocs}
can still achieve modest $\Pd$ and $\Pfa$.

%% file: content/conclusions.tex
In this paper, we have explored analog beamforming for \gls{si}-suppressed monostatic \gls{isac}.
We have identified target detection as the key performance parameter for analog beamforming, which has allowed us to formalize an optimization framework for sensing under communication constraints. In this sense, we have approached optimal \gls{isac} through parallel
\gls{tx} and \gls{rx} optimization via superiorized projections.
Finally, we have shown through simulations how this approach outperforms
a popular \gls{isac} beamforming technique.

%% file: content/ack.tex
The authors of this work acknowledge the financial support by the Federal Ministry of Education and Research of Germany (BMBF) in the programme ``Souverän. Digital. Vernetzt.'' Joint project 6G-RIC (grant numbers: 16KISK020K, 16KISK030).
Rodrigo Hernang\'{o}mez acknowledges BMBF support in the project ``6G-ICAS4Mobility'' (grant number: 16KISK235).
Zoran Utkovski acknowledges BMBF support in the project ``KOMSENS-6G'' (grant number: 16KISK121).